\documentclass[twocolumn,pra,superscriptaddress]{revtex4}
\usepackage[utf8]{inputenc}
\usepackage{amsmath}
\usepackage{amssymb}
\usepackage{bm}
\usepackage{graphicx}
\usepackage{color}

\begin{document}
\title{Effect of disorder on coherent quantum phase slips in Josephson junction chains}

\author{A. E. Svetogorov}
\affiliation{
Laboratoire de Physique et Mod\'elisation des Milieux Condens\'es,
Universit\'e de Grenoble-Alpes and CNRS,
25 rue des Martyrs, 38042 Grenoble, France}
\affiliation{Moscow Institute of Physics and Technology, 141700, Dolgoprudny, Russia}

\author{D. M. Basko}
\affiliation{
Laboratoire de Physique et Mod\'elisation des Milieux Condens\'es,
Universit\'e de Grenoble-Alpes and CNRS,
25 rue des Martyrs, 38042 Grenoble, France}

\date{\today}
\begin{abstract}
We study coherent quantum phase-slips in a Josephson junction chain, including two types of quenched disorder: random spatial modulation of the junction areas and random induced background charges. Usually, the quantum phase-slip amplitude is sensitive to the normal mode structure of superconducting phase
oscillations in the ring  (Mooij-Sch\"on modes, which are all localized by the area disorder). However, we show that the modes' contribution to the  disorder-induced phase-slip action fluctuations is small, and the fluctuations of the action on different junctions are mainly determined by the local junction parameters. We study the statistics of the total QPS amplitude on the chain and show that it can be non-Gaussian for not sufficiently long chains.
\end{abstract}

\maketitle

\section{Introduction}
One-dimensional superconductivity has been studied both theoretically \cite{Frohlich, Ferrell, Bychkov} and experimentally \cite{Giordano, Devoret} for a long time. Structures such as one-dimensional superconducting wires and Josephson junction chains are of great interest as they can be used as elements of different superconducting circuits \cite{Jung2014}. Of crucial importance in one-dimensional superconductivity is the phenomenon of phase slips, which give rise to resistance below the critical temperature $T_c$ and drive the superconductor-insulator transition \cite{Bradley1984}. Here we consider coherent quantum phase slips (QPS), which correspond to a change in the phase difference along the superconductor by $2\pi$ via quantum-mechanical tunneling without dissipation and lift ground-state degeneracy. This is a fundamental issue as it corresponds to a quantum phenomenon on macroscopic scales (length of the superconductor). Moreover, coherent quantum phase slips can be potentially used in creating phase-slip qubits \cite{Mooij2005, Manachuryan2009} or to realize a fundamental current standard in quantum metrology \cite{Flowers2004, Mooij2006, Guichard2010}.

We are interested in the regime when the phase tunneling can be described quasiclassically \cite{Matveev2002, Rastelli2013}, then each QPS corresponds to a classical imaginary-time trajectory. For a Josephson junction chain this trajectory consists of fast phase winding by $2\pi$ on one of the junctions, which gives a local contribution to the QPS action, and slow phase readjustment in the rest of the chain. This readjustment is governed by gapless Mooij-Sch\"on modes~\cite{Dahm1968, Kulik1974, Mooij1985, Camarota2001, Pop2011, Masluk2012}, which can be seen as the environment for the QPS; they produce the so-called hydrodynamic contribution to the QPS action \cite{Rastelli2013}. 

The Mooij-Sch\"on modes are sensitive to spatial variations of junction parameters, which may affect the environment contribution to the QPS action. Indeed, for a periodic spatial modulation of the chain parameters, this environment contribution was shown to be significantly modified~\cite{Svetogorov2018}. In this article we study the QPS in a disordered chain. The effect of disorder on Mooij-Sch\"on modes is quite dramatic: all modes become localized \cite{Basko2013}. We want to study how this affects the QPS.

 We consider two types of disorder: random spatial variation of the junction area and random induced charges (which can arise from random gate voltages or electronic density modulations). The effect of the latter on the QPS amplitude was studied in~\cite{Matveev2002, Ivanov2001}; it was shown that the individual QPS amplitudes on different junctions should be added with random phases, which changes the scaling of the total amplitude $W$ with the junctions number $N$ from $W\propto N$ to $W\propto\sqrt{N}$. The superconductor-insulator transition in the presence of random charges was studied in \cite{Bard2017}. However, the random charges do not modify the Mooij-Sch\"on modes. 
 
The low-frequency properties of Josephson junction chains are analogous to those of thin
superconducting wires. The environment contribution to the QPS action, determined by the Mooij-Sch\"on modes with low frequencies, is similar for wires and JJ chains. In \cite{Khlebnikov2005} the effect of random local QPS phases was adressed, and in \cite{Pai2008} randomness in the local QPS core action due to spatial variations of the wire crossection was shown to increase the wire resistivity, but the change in the Mooij-Sch\"on modes spatial structure has not been taken into account.

In this article we study the effect of disorder on both the local and the environment contributions to the QPS action. We find that the effect on the environment contribution is weaker than on the local one, and thus the localization of the Mooij-Sch\"on modes does not significantly affect the QPS amplitude. The QPS amplitude in a disordered chain is a random quantity, whose statistics is determined by the fluctuations of the local term in the QPS action. We study this statistics and we show that it can be non-Gaussian if the chain is not sufficiently long.

The structure of the article is the following. First, we introduce the model, state the problem and briefly sketch the known facts about QPS in spatially homogeneous Josephson junction chains in Sec. \ref{sec:statement}. Then, in Sec.~\ref{sec:inhomogeneous}, we propose methods to deal with disorder and discuss different regimes, depending on the type of disorder, its strength and length of the chain. In Sec. \ref{sec:wires} we discuss applicability of our results to superconducting wires. Some technical details are given in two appendices.

\section{Statement of the problem}\label{sec:statement}

\begin{figure}
\includegraphics[scale=0.33]{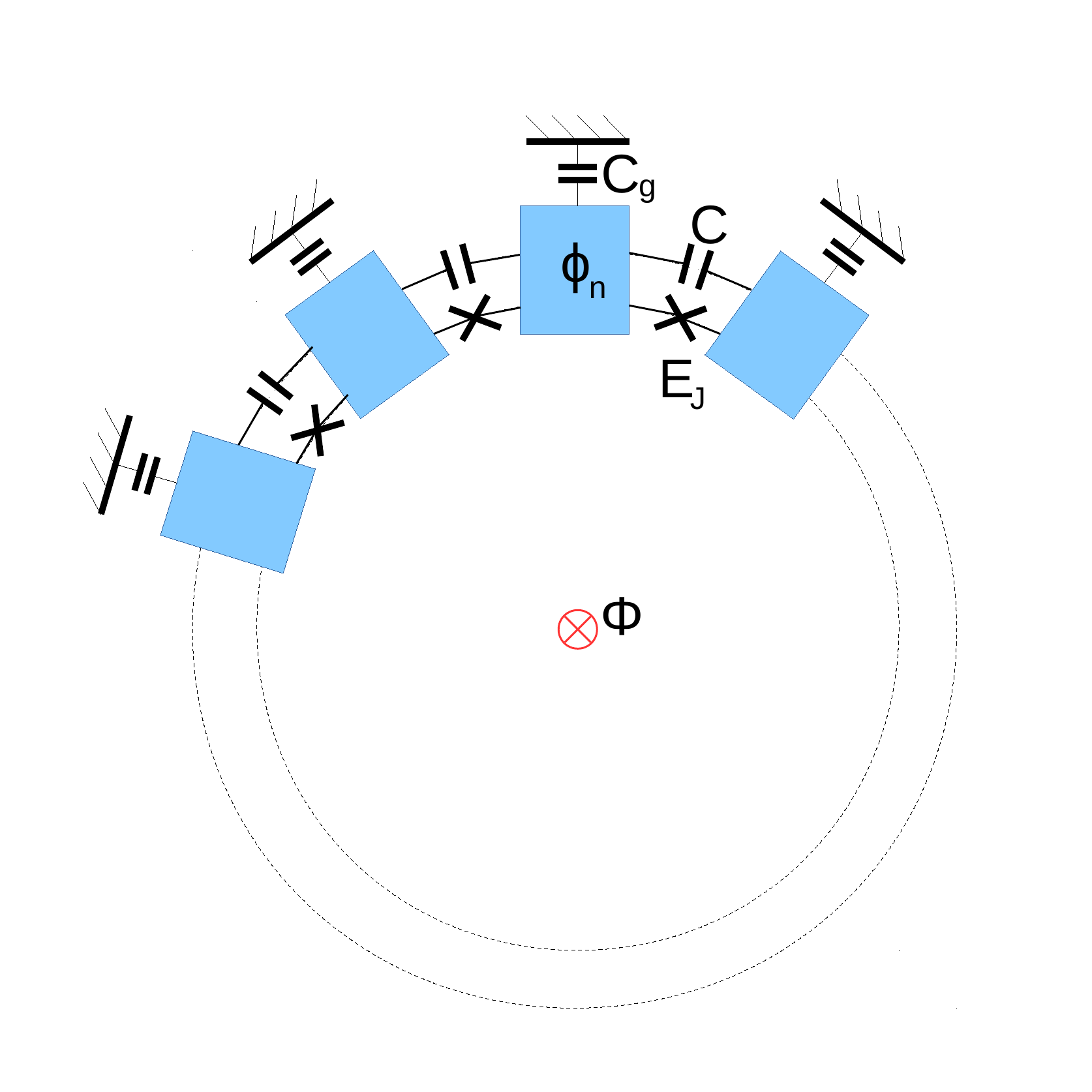}
\caption{\label{fig:JJchain}
A schematic representation of a superconducting ring
threaded by a magnetic flux $\Phi$ and containing $N$ Josephson
junctions with a capacitance $C$ between the neighbouring islands and a capacitance $C_g$ to the ground. $E_J$ is the Josephson energy. $\phi_n$ is the condensate phase of the $n$th superconducting island.}
\end{figure}

We consider a chain of $N$ Josephson junctions closed in a ring, pierced by a magnetic flux (Fig. \ref{fig:JJchain}). The superconducting islands are labeled by an integer $n$, the dynamical variables are the phases $\phi_n\left(\tau\right)$, where $\tau$ is the imaginary time. The island $n=0$ is identified with the island $n=N$, so that $\phi_0=\phi_{N}$. We describe the system by the Euclidean action~\cite{Fazio} (setting $\hbar=1$ throughout the paper):
\begin{multline}\label{eq:SJJ}
S=\int\sum_{n=0}^{N-1}\Big[ \frac{C_{\mathrm{g},n}}{8e^{2}}\,\left(\dot{\phi}_n-i\frac{2e}{C_{\mathrm{g},n}}q_n\right)^2+\\
+\frac{C_{n}}{8e^{2}}
\left(\dot{\phi}_{n+1}-\dot{\phi}_{n}\right)^2-\\
-E_{J,n}\cos\left(\phi_{n+1}-\phi_{n}+\frac{\Phi}{N}\right)\Big]{d}\tau,
\end{multline}
where $\dot\phi_n\equiv\partial\phi_n/\partial\tau$, $E_{J,n}$ and $C_{n}$ are the Josephson energy and the capacitance of the junction between neighbouring islands $n$ and $n+1$ respectively, while $C_{\mathrm{g},n}$ is the capacitance between island $n$ and a nearby ground plane. $q_n$ are the induced charges on the islands in units of the Cooper-pair charge $-2e$. $\Phi$ is the magnetic flux in units of the superconducting flux quantum divided by $2\pi$ (one flux quantum piercing the ring corresponds to $\Phi=2\pi$). It is convenient to introduce energy scales corresponding to the capacitances: 
\begin{equation}\label{eq:assumption}
E_{c,n}=\frac{e^2}{2C_n},\quad E_{g,n}=\frac{e^2}{2C_{g,n}}
\end{equation}
Typically in experiments $C_{g,n}\ll C_n$ \cite{Masluk2012, Ergul2013, Weissl2015}. We assume that 
\begin{equation}
E_J\gtrsim E_g\gg E_c,
\end{equation}
which ensures that the phase slips are rare and the chain remains superconducting for large $N$ \cite{Bradley1984, Korshunov1989, Matveev2002, Rastelli2013}. This limit is realistic and was implemented in recent experiments; for example, one of the samples in Ref. \cite{Ergul2013} had $E_J/E_c\approx90$, $E_g/E_c\approx60$.

We consider disorder in the system due to two mechanisms: random variations in the junction areas and random induced charges on the islands. Since both the Josephson energy and the capacitance of a junction are proportional to the junction area, we can represent $E_{J,n}$ and $E_{c,n}$ as
\begin{equation}\label{eq:modulation}
E_{c,n}=\frac{E_c}{1+\eta_n},\quad E_{J,n}=E_J\left(1+\eta_n\right),
\end{equation}
where $E_c$ and $E_J$ are the corresponding values of the junction parameters for a homogeneous chain and $\eta_n\ll1$ is the relative junction area variation. We consider $\eta_n$ to be independent random Gaussian with zero average and dispersion parametrized as:
\begin{equation}
\left\langle\eta^2_n\right\rangle=\frac{E_c}{8E_J}\,\sigma^2\ll1,
\end{equation} 
where $\sigma^2$ is the dispersion of the single QPS action [defined later, see Eq. (\ref{eq:sigma})]. 
As for the induced charges~$q_n$, we focus on two limiting cases: (i) all charges $q_n=0$ and (ii) charges are strongly random with the dispersion $\left\langle q_n^2\right\rangle\gg1$.

\begin{figure}
\includegraphics[width=0.45\textwidth]{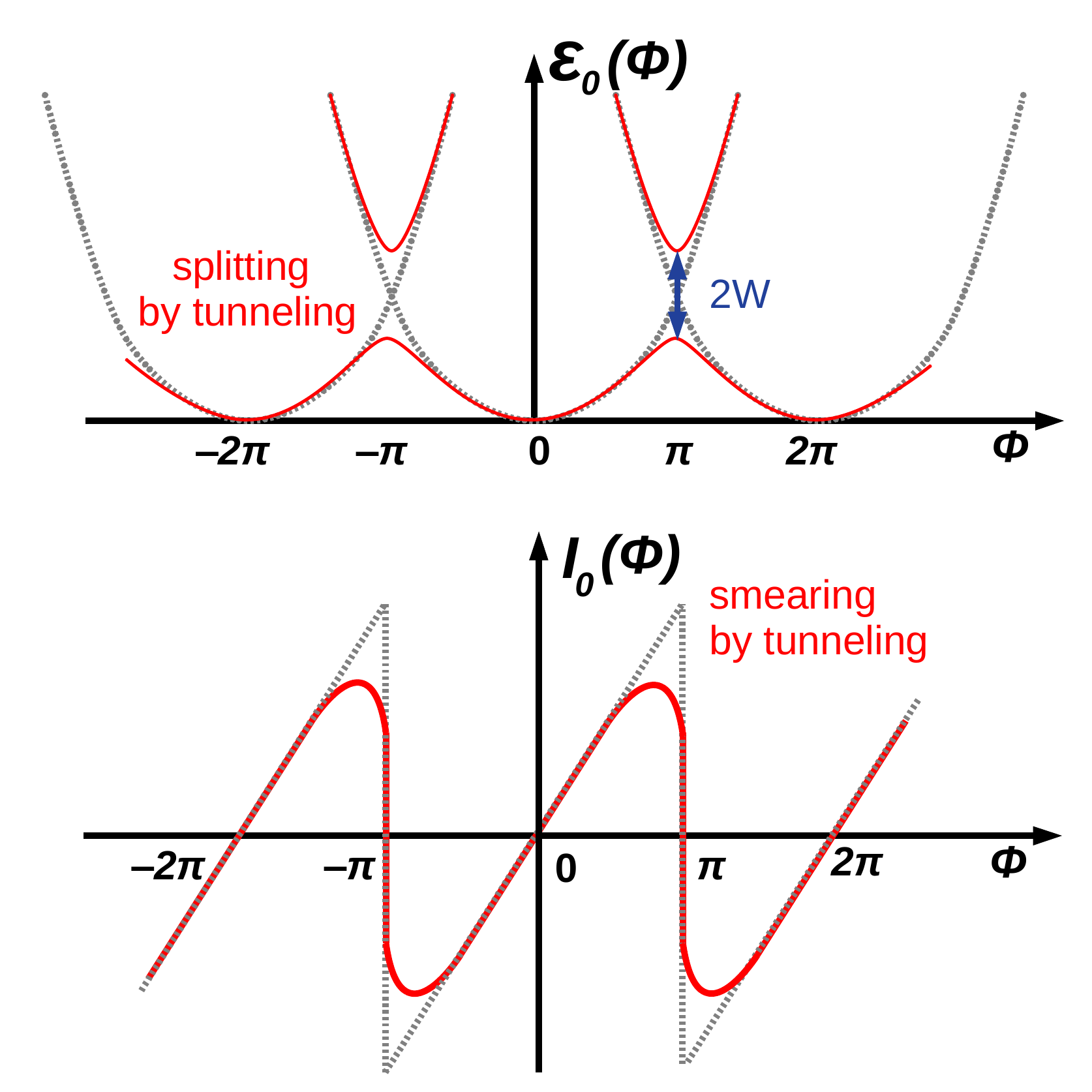}
\caption{\label{fig:GroundState}
A schematic representation of flux dependence for the ground state energy (upper panel) and persistent current (lower panel). The grey dotted lines correspond to the purely classical approximation [i. e., neglecting the kinetic terms in the action (\ref{eq:SJJ})] for the static configurations $\phi_n=2\pi n/N$, $\phi_n=0$ and $\phi_n=-2\pi n/N$ (the left, middle and right parabolas on the upper panel, respectively). The red solid lines correspond to the exact ground state including the effect of the quantum tunneling.}
\end{figure}

We study the QPS amplitude which determines the quantum tunneling splitting $2W$ between two degenerate classical ground states at $\Phi=\pi$ and the smearing of the sawtooth $\Phi$ dependence of the ground state persistent current $I_0\left(\Phi\right)\propto \partial\mathcal{E}_0(\Phi)/\partial\Phi$, where $\mathcal{E}_0(\Phi)$ is the ground state energy (Fig. \ref{fig:GroundState}). Under assumption (\ref{eq:assumption}), the QPS events are rare, then the QPS amplitude $W$ can be presented as a coherent sum of partial amplitudes of QPSs centered on different junctions:
\begin{equation}\label{eq:amplitude}
W=\sum_{n=0}^{N-1}\Omega_n e^{-S_n-i\theta_n}.
\end{equation}
Here $S_n$ is the phase action on the classical instanton trajectory, connecting the two degenerate static phase configurations which correspond to the classical ground states at $\Phi=\pi$, $\theta_n=2\pi(q_0+\ldots+q_n)$ is the random phase due to the induced charges and $\Omega_n$ is the prefactor coming from the Gaussian integration over the fluctuations around the classical trajectory.

The instanton trajectory corresponding to the $n$th term in Eq. (\ref{eq:amplitude}) involves (i) winding of the phase difference $\phi_{n+1}-\phi_n$ within the whole range $2\pi$, and (ii) phase readjustment on the rest of the junctions, where all the phase differences remain small, at most $\sim1/\ell_s\equiv\sqrt{C_g/C}\ll1$.  This readjustment is governed by the Mooij-Sch\"on modes, which play the role of the environment for the slipping junction.

Integrating out the environment degrees of freedom by the standard procedure \cite{Weiss}, one obtains the effective action for the slipping junction phase difference $\vartheta(\tau)$ (see Ref. \cite{Rastelli2013} for the spatially homogeneous case, Ref. \cite{Svetogorov2018} for the general inhomogeneous case):
\begin{align}\label{eq:instaction}
S_n[\vartheta]={}&{}\int \left\{\frac{1}{16\,E_{c,n}}\left[\frac{d\vartheta(\tau)}{d\tau}\right]^2+E_{J,n}\left[1+\cos\vartheta(\tau)\right]\right\}d\tau {}\nonumber\\
{}&{}+\frac{1}2\int
\widetilde{K}_n(\tau-\tau')\,\vartheta(\tau)\,\vartheta(\tau')\,d\tau\,d\tau'.
\end{align}
The first two terms in this action correspond to the slipping junction, while the last term, which results from integrating out the phases on the rest of the junctions, represents the action of the phase readjustment. The kernel $\widetilde{K}_n(\tau-\tau^{\prime})$ can be related to the chain impedance $Z_n\left(i\omega\right)$ at complex frequencies \cite{Vanevic2012, Svetogorov2018}:
\begin{equation}
\widetilde{K}_n(\tau-\tau')=\intop_{-\infty}^{\infty}\frac{|\omega|\,e^{-i\omega(\tau-\tau')}}{4e^2Z_n(i|\omega|)}\frac{d\omega}{2\pi}.
\end{equation}
Here $Z_n$ is the impedance of the open chain, that is the original chain open between $n$ and $n+1$ islands, which represents the physical environment for the slipping junction. $Z_n$ is determined by the Mooij-Sch\"on modes of the open chain.

The instanton trajectory $\vartheta(\tau)$  goes from $\vartheta=\pi$ to $\vartheta=-\pi$. For a homogeneous chain in the limit $C/C_{g}\gg1$ it is conveniently represented in the Fourier space \cite{Matveev2002, Svetogorov2018}
\begin{equation}\label{eq:vartheta}
\vartheta\left(\omega\right)=\frac{2\pi}{i\omega\cosh\left(\frac{\pi\omega}{4\sqrt{2E_{J}E_{C}}}\right)}.
\end{equation}
The QPS action (\ref{eq:instaction}), evaluated on this trajectory, is given by~\cite{Hekking, Matveev2002, Rastelli2013, Svetogorov2018} (for homogeneous chain we omit all the $n$ indexes) 
\begin{equation}\label{eq:homogeneous}
S_{\mathrm{hom}}=\sqrt{\frac{8E_{J}}{E_{c}}}+\sqrt{\frac{\pi^2}{8}\frac{E_{J}}{E_{g}}}\left[\ln\frac{N}{\ell_{s}}-2.43+O\left(1/\ell_s\right)\right].
\end{equation}
The first term corresponds to the slipping junction, while the rest is determined by the environment. 
The prefactor $\Omega$ is estimated to be \cite{Matveev2002}
\begin{equation}
\Omega=\frac{4}{\sqrt{\pi}}
\left(8E_{J}^3E_{c}\right)^{1/4}.
\end{equation}


\section{Spatially inhomogeneous loop: correction to the QPS amplitude}\label{sec:inhomogeneous}
\subsection{Fluctuations of the QPS action} 
We consider a JJ chain with weak relative disorder, $\langle\eta_n^2\rangle\ll1$, which produces small relative corrections to the action $S_n$ and the prefactor $\Omega_n$. While the latter results in a small relative correction to the QPS amplitude~$W$, the correction to the action, $\delta S_n$, even though small compared to $S_n$, can still be large compared to unity, since $S_n$ itself is large. As $\delta S_n$ stands in the exponent, it may significantly modify $W$. Therefore, in the following we focus on the statistics of $\delta S_n$, calculating it to the linear order in $\eta_n$. For this we can use the unperturbed expression (\ref{eq:vartheta}) for $\vartheta$ in Eq.~(\ref{eq:instaction}), because it was derived from the condition $\delta S/\delta\vartheta=0$. Then the correction to the action is:
\begin{multline}\label{eq:linear_correction}
\delta S_n=\int\left[\frac{\eta_n}{16E_c}\left(\frac{d\vartheta}{d\tau}\right)^2+\eta_n E_{J}\left(1+\cos\vartheta\right)\right]d\tau\\
+\frac{1}{2}\int\,
\delta{\widetilde{K}_n}(\omega)\,\left|\vartheta(\omega)\right|^2\frac{d\omega}{2\pi}\equiv\delta S_{n,loc}+\delta S_{n,env}.
\end{multline}

As we assume the parameters to be Gaussian distributed around the average values, the average correction to the action is zero. The quadratic fluctuations of the action are determined (i) by the variation of the slipping junction area, which in turn determines $\delta S_{n,loc}$, the first two terms in Eq.~(\ref{eq:linear_correction}), and (ii) by the correlator $\langle\delta\widetilde{K}_n\left(\omega\right)\delta\widetilde{K}_n\left(\omega^{\prime}\right)\rangle$, corresponding to the variation in the impedance of the rest of the chain, which governs $\delta S_{n,env}$, the last term in Eq.~(\ref{eq:linear_correction}). Calculation of the correlator is fully analogous to that of impedance fluctuations at real frequencies \cite{Basko2013}: using the recurrence relation for the impedance as the chain length is increased by one, one arrives at a Langevin-like equation (see Appendix \ref{app:impedance} for details). At low frequencies $\omega,\,\omega^{\prime}\ll\sqrt{8E_JE_c}$, the result is
\begin{equation}\label{eq:correlator}
\left\langle\delta\widetilde{K}\left(\omega\right)\delta\widetilde{K}\left(\omega^{\prime}\right)\right\rangle=\frac{\sqrt{2E_J}}{32E_g^{3/2}}\,\frac{\left|\omega\right|^2\left|\omega^{\prime}\right|^2}{\left|\omega\right|+\left|\omega^{\prime}\right|}\,\frac{\langle\eta^2\rangle}{2}.
\end{equation}
We are interested in the low-frequency limit of $\langle\delta\widetilde{K}\left(\omega\right)\delta\widetilde{K}\left(\omega^{\prime}\right)\rangle$ because the integrand in Eq.~(\ref{eq:linear_correction}) is quickly suppressed at $\omega>\sqrt{8E_JE_c}$ due to the frequency dependence of $\vartheta(\omega)$, Eq.~(\ref{eq:vartheta}). From this we can estimate 
\begin{equation}
\langle\delta S_{env}^2\rangle\sim\langle\eta^2\rangle\frac{\ell_s\sqrt{E_JE_c}}{E_g^2}\,\intop _0^{\sim\sqrt{E_JE_c}}\frac{d\omega\, d\omega'}{\omega+\omega^{\prime}}\sim\langle\eta^2\rangle\frac{\ell_sE_JE_c}{E_g^2}.
\end{equation}
At the same time
\begin{equation}
\langle\delta S_{loc}^2\rangle\sim\langle\eta^2\rangle\frac{E_J}{E_c}\gg\langle\delta S_{env}^2\rangle,
\end{equation}
due to the condition $C_g\ll C$.

This is one of the main results of the present work: the fluctuations of the QPS action are dominated by the local values of the slipping junction parameters, while the effect of  Mooij-Sch\"on modes modification by the disorder plays a minor role. This happened because the environment contribution to the QPS amplitude is determined by the impedance at imaginary frequencies, which turns out to be weakly fluctuating. This is in striking contrast to the behaviour at real frequencies, when localization of the Mooij-Sch\"on modes by the disorder results in strong impedance fluctuations \cite{Basko2013}.

 Having established the dominant character of the local contribution to the action fluctuations, we can study the statistics of the QPS amplitude $W$ by using Eq. (\ref{eq:amplitude}) with $S_n=S_\mathrm{hom}+\delta S_n$, where $S_\mathrm{hom}$ is the action of the homogeneous chain, Eq.~(\ref{eq:homogeneous}), and $\delta S_n$ are independent Gaussian random variables: 
 
\begin{subequations}\begin{align}
&\delta S_n=\sqrt{8\frac{E_J}{E_c}}\eta_n,\\
&\langle\delta S_n\delta S_m\rangle=8\frac{E_J}{E_c}\langle\eta_n^2\rangle \delta_{nm}=\sigma^2\delta_{nm}.\label{eq:sigma}
\end{align}\end{subequations}
This problem is addressed in the following subsections.
 
\begin{figure*}
\includegraphics[width=0.4\textwidth]{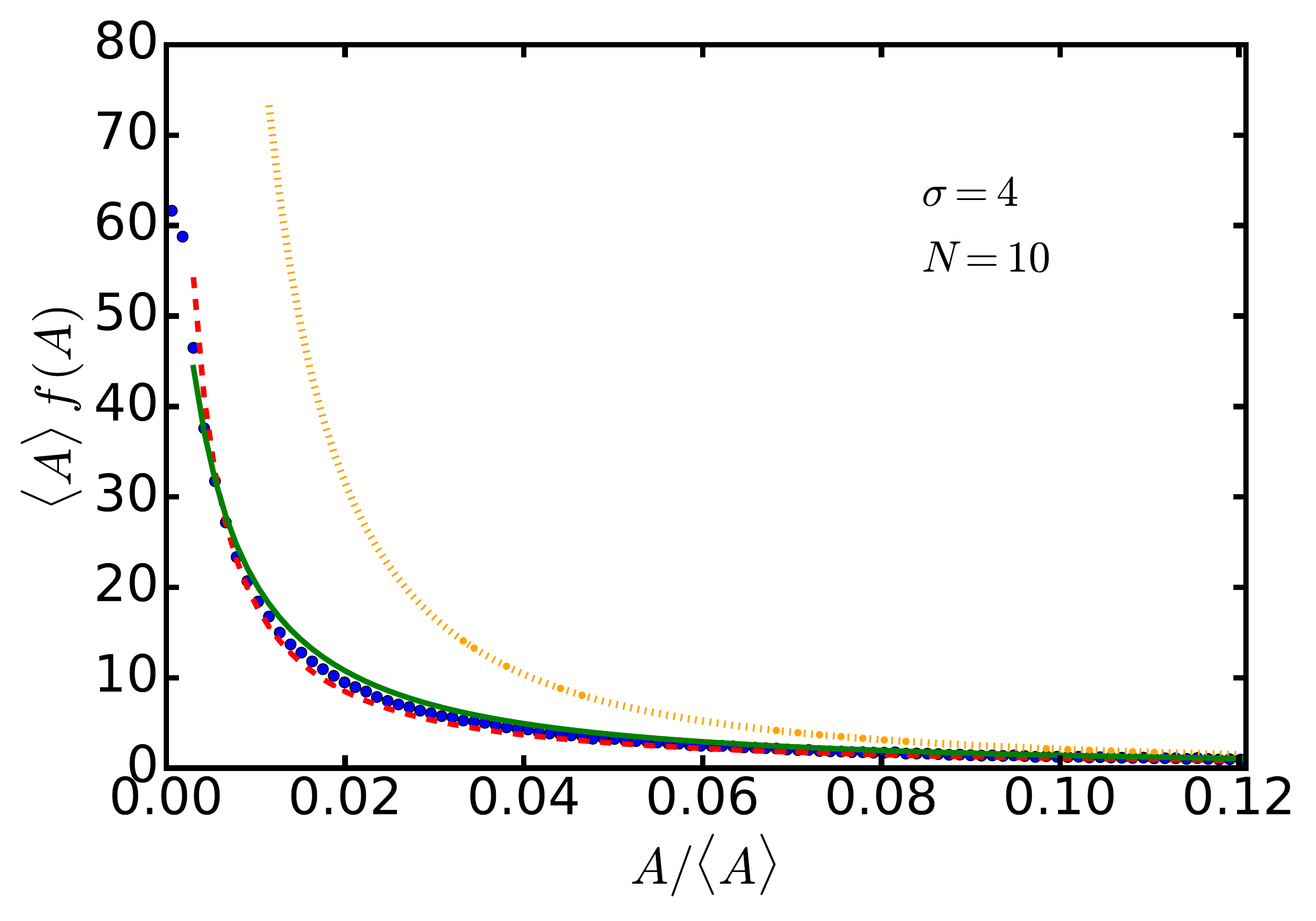}
\includegraphics[width=0.4\textwidth]{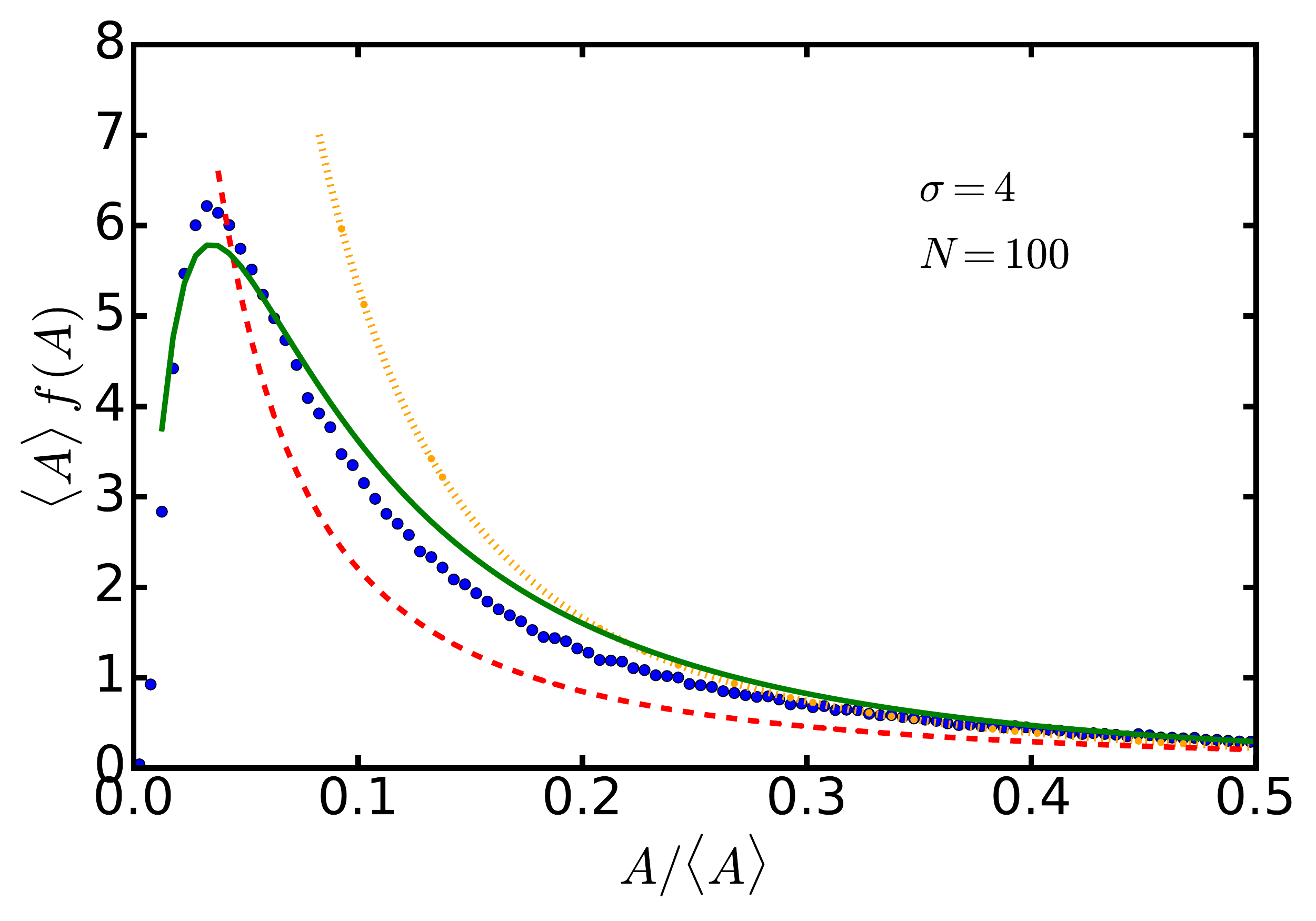}
\includegraphics[width=0.4\textwidth]{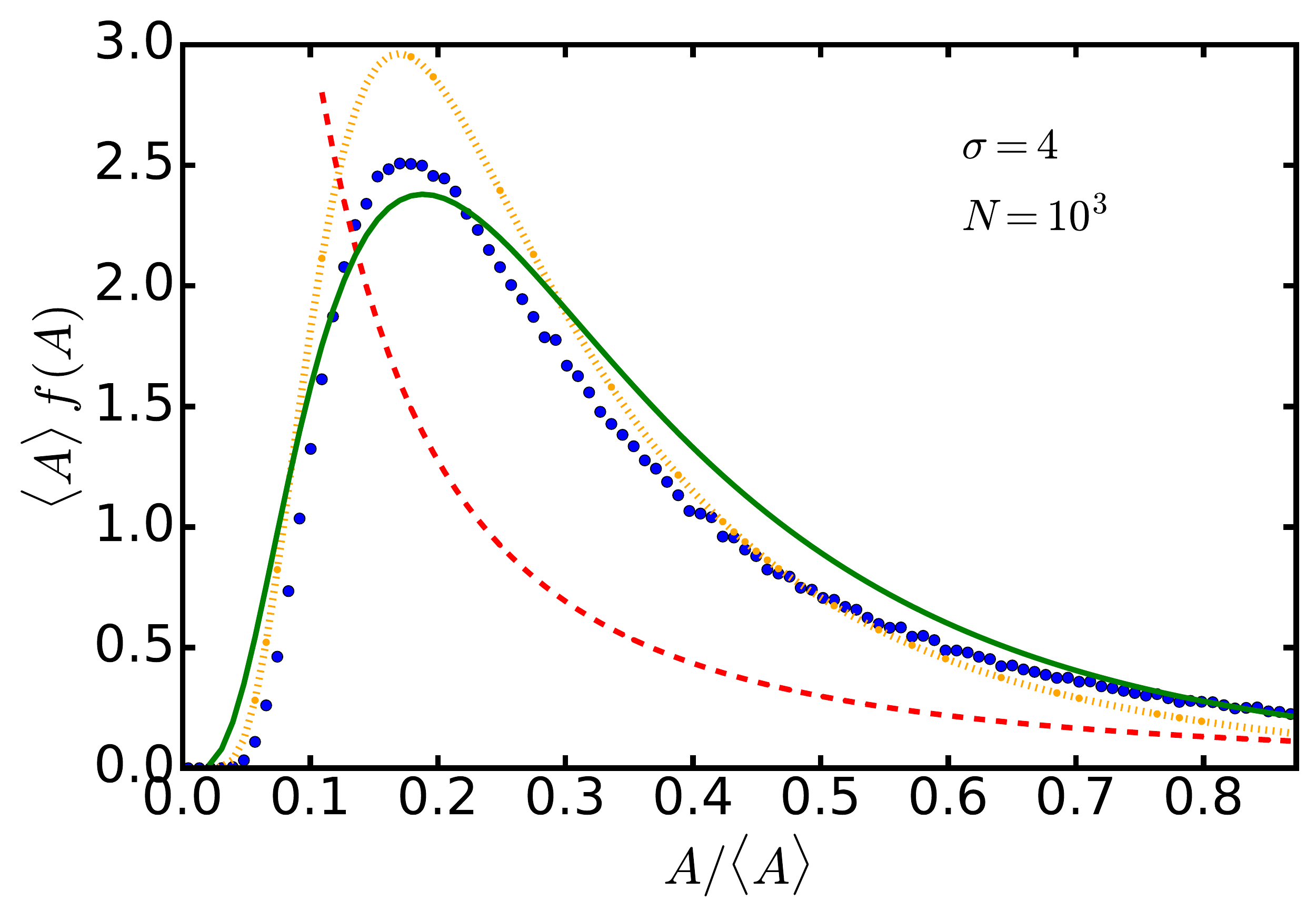}
\includegraphics[width=0.4\textwidth]{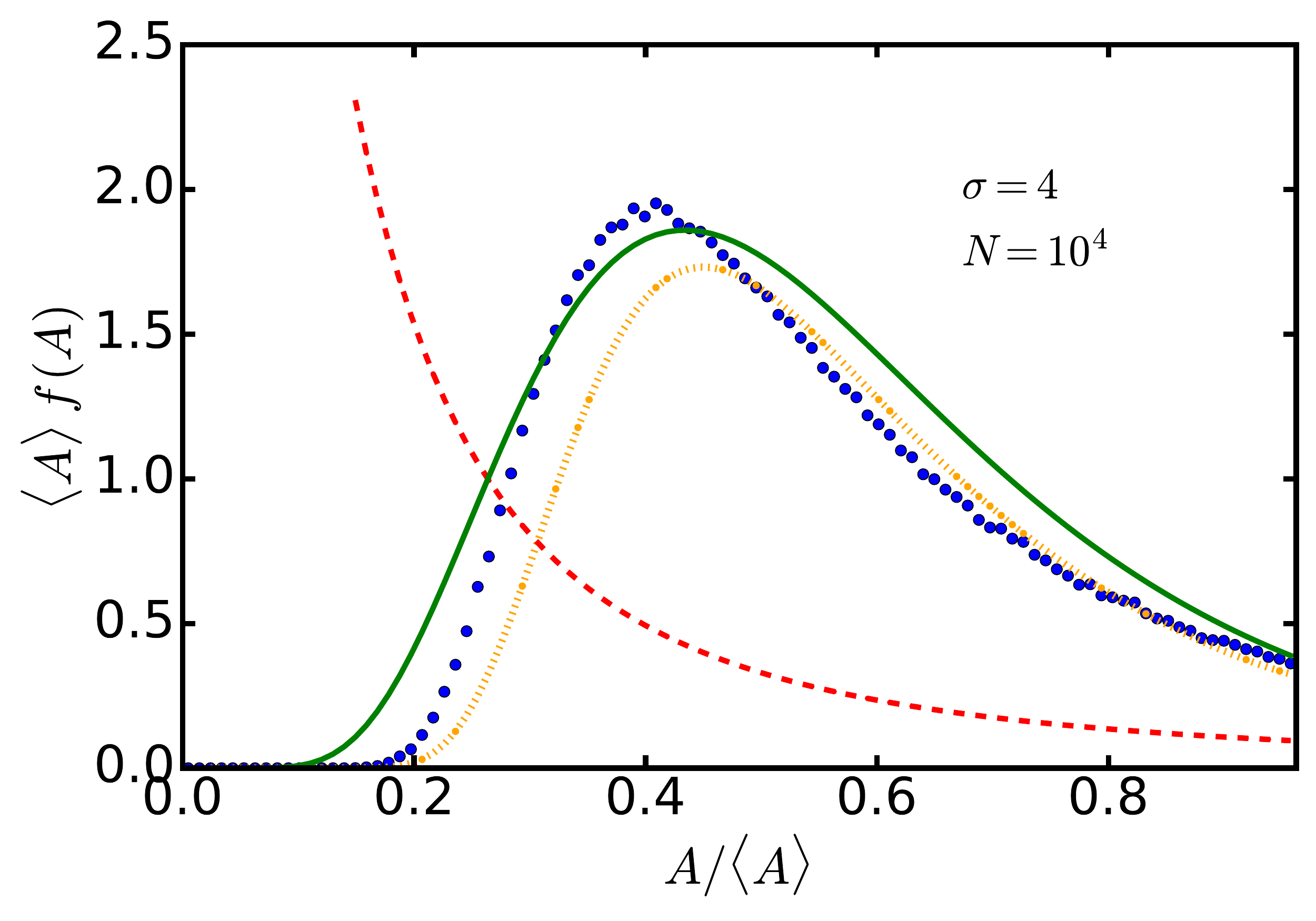}
\caption{\label{fig:Dist} Distribution  $f(A)$ in the absence of induced charges, calculated for $\sigma=4$ and different $N$ by the direct numerical sampling (blue dots), using the weakest junction approximation (\ref{eq:weakest}) (red dashed lines), the saddle-point approximation (\ref{eq:steepest}) (orange dotted lines), and the lognormal fit (solid green lines).}
\end{figure*}

\subsection{QPS amplitude distribution without random induced charges} \label{ssec:distribution}

First, we consider only the junction area variation assuming no induced charges. For long chains we can use the central limit theorem resulting in the Gaussian distribution with the average amplitude and dispersion
\begin{subequations}\begin{align}
&\langle W\rangle=\Omega\,e^{-S_\mathrm{hom}}N\,e^{\sigma^2/2},\\
&\sqrt{\langle W^2\rangle-\langle W\rangle^2}=\Omega\,e^{-S_\mathrm{hom}}\sqrt{N(e^{2\sigma^2}-e^{\sigma^2})}.
\end{align}\end{subequations}
 The central limit theorem is valid when the dispersion is much smaller than the average, that is $N\gg e^{\sigma^2}-1$. However, even for small relative area fluctuations $\langle\eta_n^2\rangle\ll1$, it is quite possible that $\sigma^2\gtrsim1$. Indeed, taking the above cited parameters of experiment \cite{Ergul2013}, $E_J/E_c\approx90$, and assuming $\langle\eta_n^2\rangle=10^{-2}$, we obtain $\sigma^2\approx7$. Then the central limit theorem applies only for exponentially large~$N$. 
 
 For $\sigma>1$ and insufficiently large $N$, the distribution can be far from Gaussian; it develops a long asymmetric tail for large $W$. In fact, this problem is known since long ago in many different areas, such as communications \cite{Marlow1967, Beaulieu1995}, optics \cite{Mitchell1968}, transport in disordered systems \cite{Raikh1987}, finances \cite{Dufresne2004}, yet no general analytical expression for the resulting distribution is available. Sometimes the resulting distribution can be approximated by a lognormal one \cite{Mitchell1968, Dufresne2004, Beaulieu1995}. Below we revisit this problem for $\sigma^2\gtrsim1$ and give some analytical expressions valid in different regimes [Eqs. (\ref{eq:steepest}) and (\ref{eq:weakest})], and compare them to the results of the direct numerical sampling and its lognormal fit (Fig.~\ref{fig:Dist}).
 
To derive analytical expressions, let us represent the QPS amplitude as $W=A\Omega\,e^{-S_\mathrm{hom}}$, then the distribution function for the normalized amplitude $A$ is defined as
\begin{align}
f\left(A\right)=&\left\langle \delta\left(A-\sum_{n=1}^{N}\exp\left(-\delta S_{n}\right)\right)\right\rangle =\nonumber\\
=&\int\frac{dt}{2\pi}\,e^{itA}\left[\int\frac{dx}{\sqrt{2\pi}\sigma}e^{-\frac{x^{2}}{2\sigma^{2}}}\exp\left(-ite^{-x}\right)\right]^{N}.\label{eq:distribution}
\end{align}
The average value $\langle A\rangle=N\,e^{\sigma^2/2}$.

The $t$ integral can be calculated in the saddle-point approximation similarly to Ref. \cite{Raikh1987} (for details see Appendix \ref{app:steepest_descent}). This calculation, valid at $N\gtrsim\sigma e^{\sigma^{2}/2-\sigma}\gg1$, gives
\begin{align}\label{eq:steepest}
&f\left(A\right)\approx\frac{\sigma M^{1/2}}{Ne^{\sigma^{2}/2}}\exp\left(-Me^{\sqrt{2}\sigma Q-Q^{2}}+\frac{Q^{2}+\sqrt{2}\sigma Q}{2}\right),\\
 &Q\equiv\mathrm{erfc}^{-1}\left(\frac{2A}{Ne^{\sigma^{2}/2}}\right),\quad
M\equiv\frac{Ne^{-\sigma^{2}/2}}{\sqrt{2\pi}\sigma e}.\nonumber
\end{align}

\begin{figure}
\includegraphics[width=0.48\textwidth]{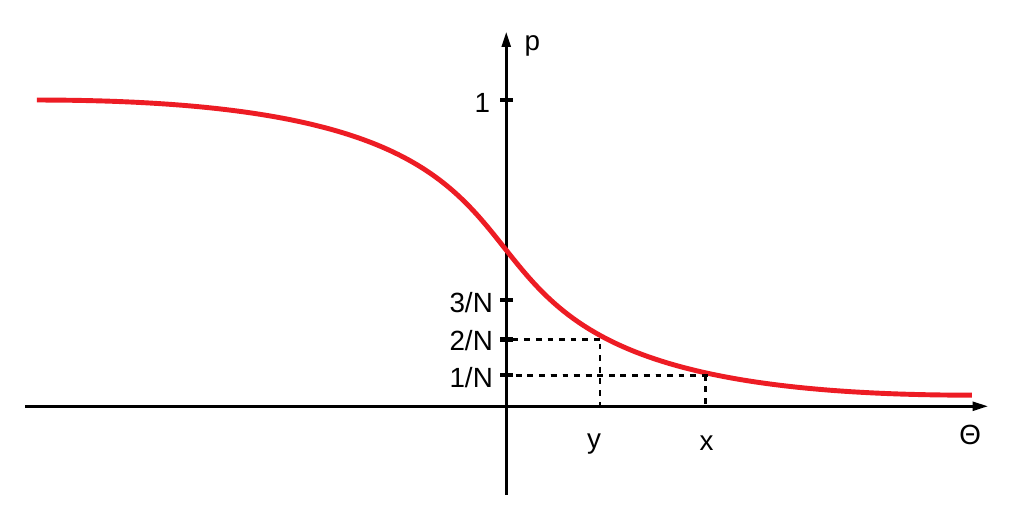}
\caption{\label{fig:Erfc} Cumulative probability distribution of $\delta S_n$ and estimates of  the two smallest $\delta S_n$ for a typical sample.}
\end{figure}

Another analytically tractable regime is when the whole sum is determined by a single term, corresponding to the junction with the highest QPS amplitude (the weakest junction).  The probability of having one junction with $x<\delta S_n<x+dx$ and the rest of the junctions with $\delta S_n<x$ is 
\begin{equation}
p\left(x\right)\,dx=\left(\,\,\intop_{-\infty}^{x}\frac{dx}{\sqrt{2\pi}\sigma}e^{-\frac{x^{2}}{2\sigma^{2}}}\right)^{N-1}\frac{N}{\sqrt{2\pi}\sigma}e^{-\frac{x^{2}}{2\sigma^{2}}}\,dx,
\end{equation}
where $N$ in the last factor corresponds to the fact that the junction with the highest amplitude can be any of the $N$ junctions. Then for the distribution we have
\begin{multline}\label{eq:weakest}
f\left(A\right)=\int\delta\left(A-e^{-x}\right)p\left(x\right)dx=\\
=\frac{N}{\sqrt{2\pi}\sigma A}\exp\left[-\frac{\left(N-1\right)}{2}\mathrm{erfc}\left(\frac{\ln A}{\sqrt{2}\sigma}\right)-\frac{\ln^{2}A}{2\sigma^{2}}\right].
\end{multline}
The weakest junction approximation is valid when the amplitude on the weakest junction, $\exp(-\mathrm{min}\{\delta S_n\})$, is sufficiently larger than the sum of the amplitudes on the rest of the junctions, which can be estimated from above as $(N-1)\exp(-\mathrm{min}^\prime\{\delta S_n\})$, where $\mathrm{min}^\prime\{\delta S_n\}$ denotes the second smallest of $\{\delta S_n\}$. To estimate the typical values of the two smallest $\delta S_n$, we recall the standard procedure for sampling the Gaussian distribution: from a sample of $N$ numbers $\{x_n\}$, uniformly distributed between $0$ and $1$, one obtains a sample of the Gaussian $\{\delta S_n\}$ by taking the inverse of the cumulative probability distribution function (see Fig.~\ref{fig:Erfc}). In a typical sample, $\mathrm{min}\{x_n\}\sim1/N$ and $\mathrm{min}'\{x_n\}-\mathrm{min}\{x_n\}\sim1/N$, so we estimate
\begin{equation}
\frac{1}{2}\,\mathrm{erfc}\left(\frac{\mathrm{min}\{\delta S_n\}}{\sqrt{2}\sigma}\right)=\frac{1}{N},\quad \frac{1}{2}\,\mathrm{erfc}\left(\frac{\mathrm{min}'\{\delta S_n\}}{\sqrt{2}\sigma}\right)=\frac{2}{N}.
\end{equation}
This results in the validity condition 
\begin{equation}\label{eq:validity}
N\lesssim \exp\left[\left(\ln^{2}2/2\right)^{1/3}\sigma^{2/3}\right].
\end{equation}

\begin{figure*}
\includegraphics[width=0.4\textwidth]{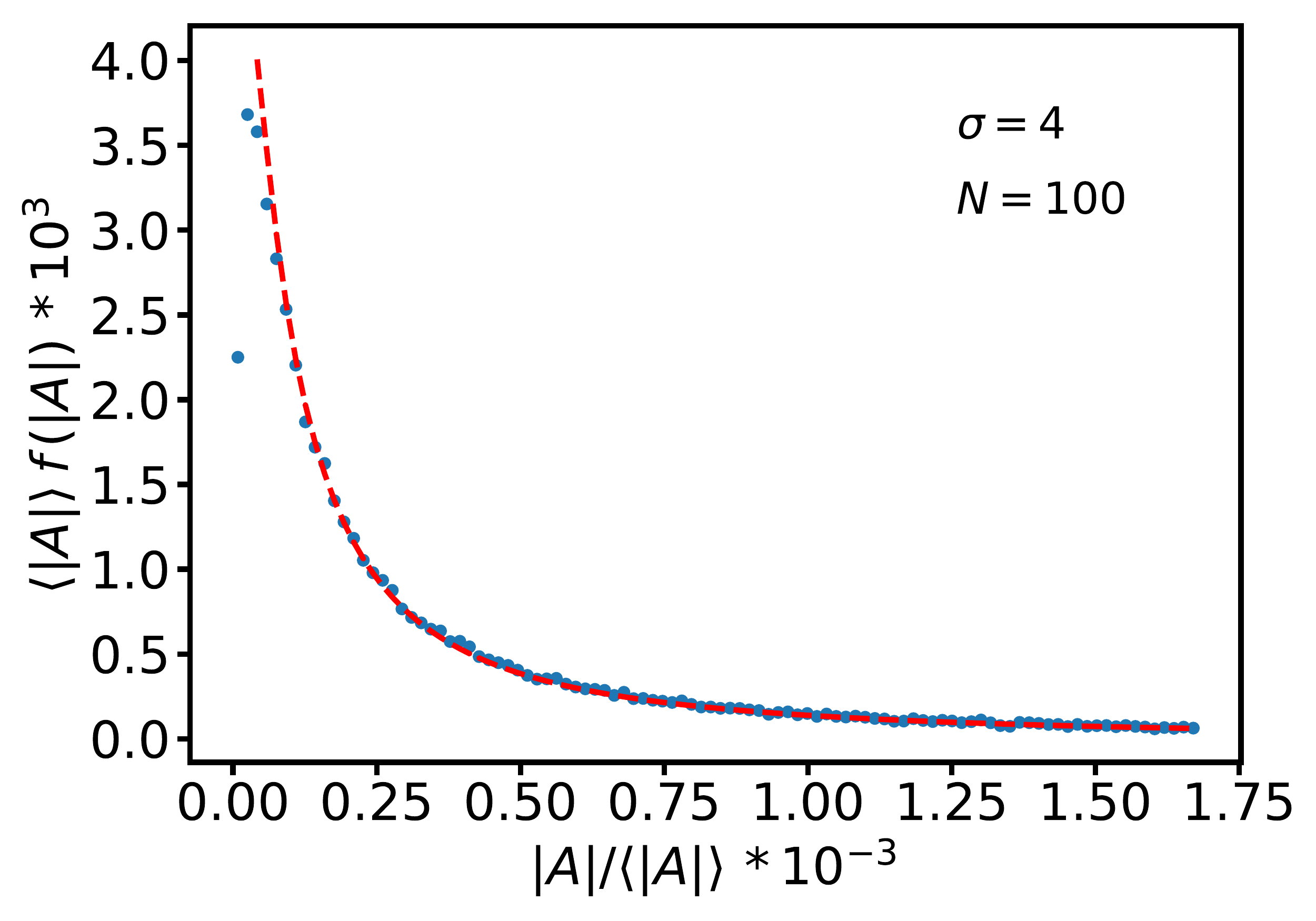}
\includegraphics[width=0.4\textwidth]{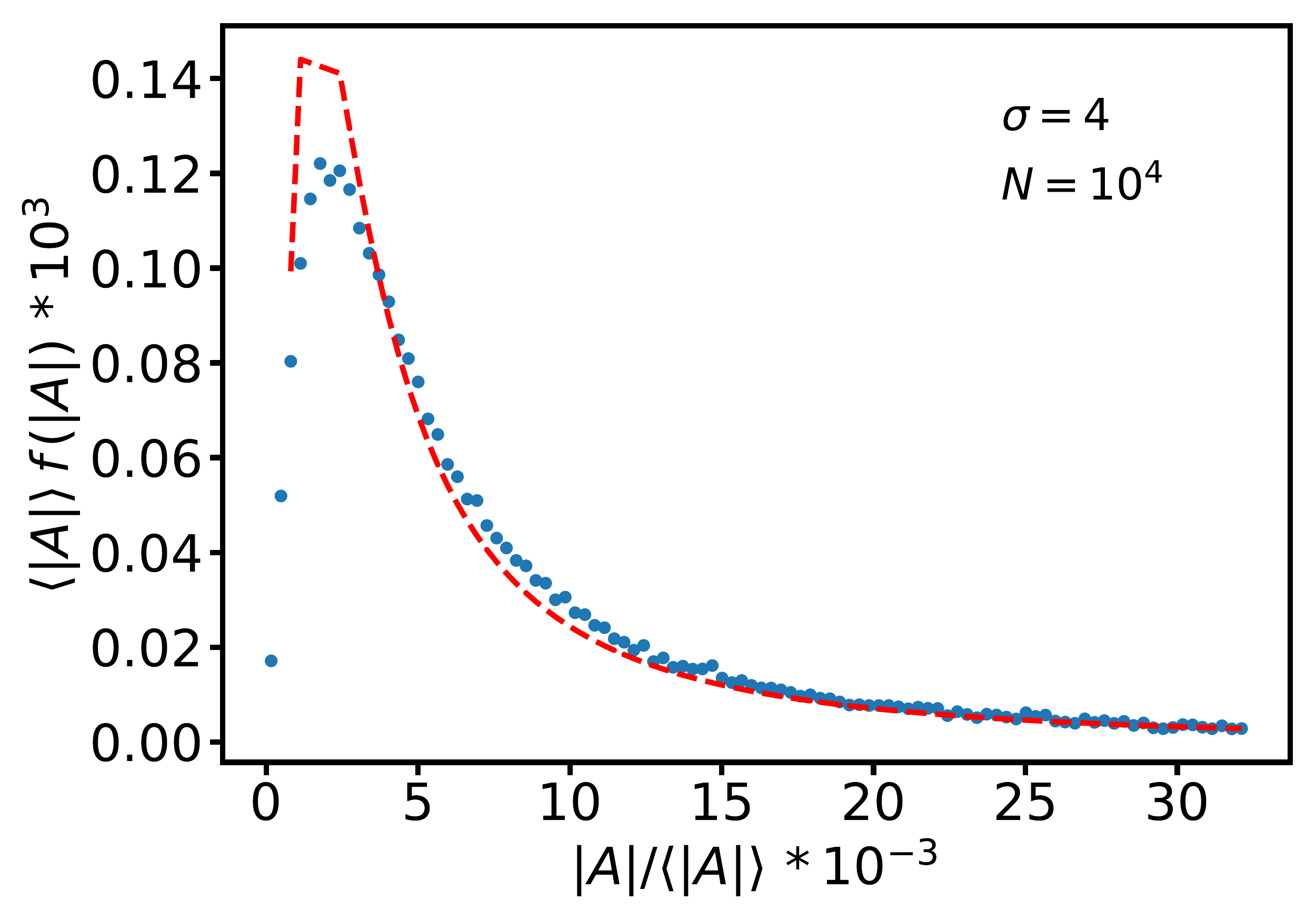}
\caption{\label{fig:Dist_with_charges}Distribution  $f(|A|)$ with random induced charges, calculated for $\sigma=4$ and different $N$ by the direct numerical sampling (blue dots) and using the weakest junction approximation (\ref{eq:weakest}) (red dashed lines). }
\end{figure*}
\subsection{QPS amplitude distribution with random induced charges}\label{ssec:induced}

If we include random induced charges, which are sufficiently strong ($|q_n|\gg1$), we obtain a random phase in the amplitude of a single QPS centered on each junction \cite{Ivanov2001, Matveev2002}, see Eq. (\ref{eq:amplitude}). Then the normalized QPS is given by 
\begin{equation}
A=\sum_{n=0}^{N-1}e^{-\delta S_n-i\theta_n}.
\end{equation}
Therefore, $A$ is complex and its average  is zero. The central limit theorem results in the complex Gaussian distribution with
\begin{equation}
\sqrt{\langle|A|^2\rangle}=\sqrt{N}\,e^{\sigma^2}.
\end{equation}
The criterion for the validity of the central limit theorem is the correspondence of the moments of $A$ to the moments of the complex Gaussian distribution, for example
\begin{equation}
\langle|A|^4\rangle-2\langle|A|^2\rangle^2\ll \langle|A|^4\rangle.
\end{equation} 
This results in the condition $N\gg (e^{4\sigma^2}-1)/2$, even more restrictive than in the real case.

In the complex case, we were unable to derive a compact expression for the distribution function corresponding to the saddle point approximation. The weakest junction approximation works when
\begin{equation}\label{eq:restric_charge}
N\lesssim \exp\left[\left(2\ln^{2}2\right)^{1/3}\sigma^{2/3}\right].
\end{equation}
 Then the distribution of $|A|$ is the same as the distribution of $A$ in Eq. (\ref{eq:weakest}). The only difference is in the restriction on the chain length $N$: the approximation is valid for a wider range of $N$ as seen from Eq. (\ref{eq:restric_charge}) and Eq. (\ref{eq:validity}), see Fig.~\ref{fig:Dist_with_charges}.

\section{Inhomogeneous superconducting wires}\label{sec:wires}

Let us discuss applicability of our results to the QPS in superconducting wires. Indeed, the low energy excitations (the Mooij-Sch\"on modes) are similar for wires and for Josephson junction chains. The low energy properties of a superconducting wire are determined by the inductance per unit length, $\cal L$, and ground capacitance per unit length, $\cal C$. We can represent a superconducting wire as Josephson junction chain with parameters $E_J$, $C_g$ and junction size $a$ by matching the Mooij-Sch\"on mode velocity and the low frequency wire admittance: 
\begin{equation}
\frac{1}{\sqrt{{\cal LC}}}=a\sqrt{8E_JE_g},\quad\sqrt{{\cal \frac{C}{L}}}=\sqrt{\frac{E_J}{8E_g}}.
\end{equation}
While for Josephson junction chains the frequency cut-off is $\sqrt{8E_JE_c}$, for wires it is given by the superconducting gap $2\Delta$.
The analog of the random spatial variation $E_{J,n}=E_J(1+\eta_n)$ would be the spatial variation ${\cal L}(x)={\cal L}/[1+\eta(x)]$, which can result from, e.g., spatial fluctuations in the wire thickness on the spatial scale exceeding the thickness itself and the superconducting coherence length $\xi$. The parameters ${\cal L}(x)$ and ${\cal C}(x)$ are already averaged over the microscopic disorder due to impurities, acting on the length scale shorter than $\xi$. Then instead of $\langle\eta_n\eta_m\rangle=D\delta_{nm}$ with $D\ll1$ we have $\langle\eta(x)\eta(x')\rangle={\cal D}\delta(x-x')$, where ${\cal D}$ has the dimensionality of length and $\delta(x-x')$ is peaked on the length $\sim\xi$. If we represent a segment of the wire of length $a\gtrsim\xi$ by a Josephson junction with $E_J=1/[(2e)^2a\cal L]$, then $D={\cal D}/a$. Thus, the weak-disorder condition is ${\cal D}\ll\xi$.

Similarly to the Josephson junction chains, the QPS action in superconducting wires can be represented as a sum of two contributions: $S_\mathrm{QPS}=S_\mathrm{loc}+S_\mathrm{env}$. The environment part of the action is also determined by the Mooij-Sch\"on modes. This enables us to use the result of appendix \ref{app:impedance} for the low-frequency admittance fluctuations:
\begin{equation}
\left\langle\delta\widetilde{K}\left(\omega\right)\delta\widetilde{K}\left(\omega^{\prime}\right)\right\rangle=\frac{{\cal C}^{3/2}{\cal L}^{-1/2}\left|\omega\right|^2\left|\omega^{\prime}\right|^2}{2(2e)^4\left(\left|\omega\right|+\left|\omega^{\prime}\right|\right)}\,\cal D.
\end{equation}
Using the estimate $\Delta^{-1}$ for the instanton duration \cite{Vanevic2012}, from Eq. (\ref{eq:linear_correction}) we obtain an estimate 
\begin{equation}
\langle\delta S_\mathrm{env}^2\rangle\sim\frac{{\cal C}\sqrt{{\cal C}/{\cal L}}\Delta{\cal D}}{(2e)^4}.
\end{equation}
 The local part of the QPS action can not be calculated precisely for superconducting wires~\cite{Golubev2001, Vanevic2012}. However, it can be estimated as \cite{Vanevic2012} $S_\mathrm{loc}\sim \frac{1}{(2e)^2{\cal L}\xi\Delta}$, which gives
\begin{equation}
\langle\delta S_\mathrm{loc}^2\rangle\sim\frac{{\cal D}/\xi}{(2e)^4{\cal L}^2\xi^2\Delta^2}.
\end{equation}
As a result, we have $\langle\delta S_\mathrm{loc}^2\rangle\gg\langle\delta S_\mathrm{env}^2\rangle$ if 
\begin{equation}\label{eq:velocity}
\xi\ll\frac{1}{\Delta\sqrt{{\cal LC}}}.
\end{equation} 
In fact, this relation usually holds for superconducting wires because the mode velocity $1/\sqrt{\cal LC}$ is sufficiently high. Indeed, $1/{\cal C}$ has two contributions: one from the quantum capacitance of the Fermi sea, and the electrostatic contribution due to Coulomb interaction. In the absence of Coulomb interaction the mode velocity would be such that both sides of Eq.~(\ref{eq:velocity}) would be of the same order. However, the Coulomb contribution is usually much stronger, so the velocity is high enough to ensure the strong inequality (\ref{eq:velocity}). The right-hand side of this inequality can be seen as an analogue of $\ell_s$ for the superconducting wires, and inequality (\ref{eq:velocity}) is an analogue of $\ell_s\gg1$.

As a result, analogously to the JJ chains, the fluctuations of the QPS actions are determined by the local values of the wire parameters in the phase-slip core of the size $\xi$.

\section{Conclusions and outlook}\label{sec:conclusion}
In conclusion, we have studied the coherent QPS in disordered Josephson junction chains. We consider two sources of disorder: random junction area variation and random induced charges on the superconducting islands. We find that the main correction to the QPS amplitude $W$ is determined not by the environment contribution to the QPS action $\delta S_\mathrm{env}$, but mostly by the local values of the slipping junction parameters. This means that the Mooij-Sch\"on modes localisation does not significantly affect the QPS amplitude. 

We have studied the statistics of the QPS amplitude $W$, which is given by the sum of individual (random) amplitudes on different junctions. For very long chains $W$ has a Gaussian distribution according to the central limit theorem. However, as the fluctuations of the QPS action can be large compared to unity, for insufficiently large $N$, the distribution of $W$ is non-Gaussian and has a long tail at large $W$. We have studied different regimes, depending on the type of disorder (with zero or strongly random induced charges), chain length and strength of the disorder.

We have also discussed the QPS in spatially disordered superconducting wires. Our estimates show that the main effect of inhomogeneity on the QPS amplitude is also due to the local parameters of the superconducting wire in the core region of the size of the superconducting coherence length.

\acknowledgements

This work was supported by the French Agence Nationale de la Recherche (ANR) under grant ANR-15-CE30-0021 ``QPSNanoWires''.

\appendix
\section{Admittance fluctuations}\label{app:impedance}
The basic idea of the approach is to study the change in the admittance $Y_N(i\omega)\equiv1/Z_N(i\omega)$ of an open chain of $N$ Josephson junctions upon addition of an extra junction $N+1$. We can write the following recurrence relation for the admittance:
\begin{equation}\label{eq:recurrence}
Y_{N+1}=\omega C_{g}+\frac{Y_{N}Y_{J}}{Y_{N}+Y_{J}},
\end{equation}
where $Y_{J}=1/(\omega L_{N+1})+\omega C_{N+1}$ is the imaginary frequency admittance of the added junction and the Josephson inductance is defined as $1/L_{N+1}=\left(2e\right)^2E_{J,N+1}$.

First, let us consider a homogeneous chain. Then the recurrence relation (\ref{eq:recurrence}) has a stationary point $Y_{\infty}$, determined by the condition
\begin{equation}
Y_{\infty}=\omega C_{g}+\frac{Y_{\infty}Y_{J}}{Y_{\infty}+Y_{J}},
\end{equation}
which gives
\begin{equation}
Y_{\infty}=\frac{\omega C_{g}}{2}+\sqrt{\frac{\omega^{2}C_{g}^{2}}{4}+\omega C_{g}Y_{J}}\approx\sqrt{\omega C_{g}Y_{J}}.
\end{equation}
The latter approximation follows from $C\gg C_g$. Focusing on small deviations from the stationary point, we introduce the new variable $X_{N}=Y_{N}-Y_{\infty}$. The linearized recurrence relation takes a simple form:
\begin{equation}
X_{N+1}=\tau\,X_{N},\quad\tau\equiv\frac{Y_{J}^{2}}{\left(Y_{\infty}+Y_{J}\right)^{2}}.
\end{equation}
Note that $1-\tau\ll1$, following from $C_g\ll C$.

Now we can include fluctuations of the chain parameters,
\begin{equation}
C_{N+1}\rightarrow C\left(1+\eta_{N+1}\right),\quad L_{N+1}\rightarrow\frac{L}{1+\eta_{N+1}},
\end{equation}
and write the linearized recurrence relation as
\begin{equation}\label{eq:linrecurrence}
X_{N+1}=\tau X_{N}+\frac{Y_{N}^{2}Y_{J}}{\left(Y_{N}+Y_{J}\right)^{2}}\,\eta_{N+1}=\tau X_{N}+\delta X_{N+1}.
\end{equation}
Using the condition $1-\tau\ll1$ we can cast this equation into a differential form:
\begin{equation}
\frac{dX_N}{dN}=-(1-\tau)X_N+\frac{Y_{N}^{2}Y_{J}}{\left(Y_{N}+Y_{J}\right)^{2}}\,\eta_{N+1},
\end{equation}
which is a Langevin equation with It\^o prescription for the multiplicative noise term \cite{VanKampen}.

So far we considered the admittance at a given frequency~$\omega$. We are interested in the correlator of admittances at two different frequencies $\omega$ and $\omega^{\prime}$. Then taking into account the fact that $\delta X_{N+1}$ and $X_{N}$ are not correlated (It\^o prescription) we can average the product of equations (\ref{eq:linrecurrence}) at different frequencies:
\begin{align}
\left\langle X_{N+1}\left(\omega\right)X_{N+1}\left(\omega^{\prime}\right)\right\rangle =\tau\left(\omega\right)\tau\left(\omega^{\prime}\right)\left\langle X_{N}\left(\omega\right)X_{N}\left(\omega^{\prime}\right)\right\rangle+\nonumber\\ +\left\langle \delta X_{N+1}\left(\omega\right)\delta X_{N+1}\left(\omega^{\prime}\right)\right\rangle ,
\end{align}
which can again be rewritten as a differential equation:
\begin{align}
&\frac{d}{dN}\left\langle X_{N}\left(\omega\right)X_{N}\left(\omega^{\prime}\right)\right\rangle=\nonumber\\ &{}=\left(\tau\left(\omega\right)\tau\left(\omega^{\prime}\right)-1\right)\left\langle X_{N}\left(\omega\right)X_{N}\left(\omega^{\prime}\right)\right\rangle+\nonumber\\ &{}+\left\langle \delta X_{N+1}\left(\omega\right)\delta X_{N+1}\left(\omega^{\prime}\right)\right\rangle .
\end{align}
As we consider long chains, we can go to the limit $N\rightarrow\infty$ and look for the stationary solution:
\begin{equation}
\left\langle X\left(\omega\right)X\left(\omega^{\prime}\right)\right\rangle =\frac{\left\langle \delta X\left(\omega\right)\delta X\left(\omega^{\prime}\right)\right\rangle }{1-\tau\left(\omega\right)\tau\left(\omega^{\prime}\right)},
\end{equation}
which is the correlator of admittance fluctuations. The correlator of the kernel fluctuations is
\begin{equation}
\left\langle\delta\widetilde{K}\left(\omega\right)\delta\widetilde{K}\left(\omega^{\prime}\right)\right\rangle=\frac{\left|\omega\right|\left|\omega^{\prime}\right|}{\left(2e\right)^4}\left\langle X\left(\omega\right)X\left(\omega^{\prime}\right)\right\rangle.
\end{equation}
Evaluating $\left\langle \delta X\left(\omega\right)\delta X\left(\omega^{\prime}\right)\right\rangle$ from the definition (\ref{eq:linrecurrence}) and collecting all factors, we obtain the following behaviour in the two limiting cases. At $\omega,\,\omega^{\prime}\ll1/\sqrt{LC}$,
\begin{equation}\label{eq:KK}
\left\langle\delta\widetilde{K}\left(\omega\right)\delta\widetilde{K}\left(\omega^{\prime}\right)\right\rangle=\frac{C_{g}^{3/2}L^{-1/2}\left|\omega\right|^2\left|\omega^{\prime}\right|^2}{2(2e)^4\left(\left|\omega\right|+\left|\omega^{\prime}\right|\right)}\,\langle\eta^2\rangle,
\end{equation}
which is Eq. (\ref{eq:correlator}). Note that the capacitance $C$ dropped out from this result; this occurs regardless of our assumption $C\gg C_g$ and is the consequence of the low frequency limit. For $\omega,\,\omega^{\prime}\gg1/\sqrt{LC}$ we have 
\begin{equation}
\left\langle\delta\widetilde{K}\left(\omega\right)\delta\widetilde{K}\left(\omega^{\prime}\right)\right\rangle=\frac{C_{g}^{3/2}C^{1/2}\left|\omega\right|^2\left|\omega^{\prime}\right|^2}{4(2e)^4}\,\langle\eta^2\rangle.
\end{equation}

\section{Saddle point approximation for the probability distribution}\label{app:steepest_descent}

Let us rotate the integration contour in Eq. (\ref{eq:distribution}) to the imaginary axis:
\begin{align}
&f\left(A\right)=\intop_{-i\infty}^{i\infty}\exp\left[zA-N\,I(z)\right]\frac{dz}{2\pi i}, \label{eq:doubleintegral}\\
&I(z)\equiv-\ln\left[\int\frac{dx}{\sqrt{2\pi}\sigma}e^{-\frac{x^{2}}{2\sigma^{2}}}\exp\left(-ze^{-x}\right)\right].
\end{align}
In the saddle-point approximation, we have
\begin{equation}
f\left(A\right)\approx\sqrt{\frac{1}{2\pi NI^{\prime\prime}\left(z_{s}\right)}}\exp\left[z_sA+NI\left(z_{s}\right)\right],
\end{equation}
where $I'(z)=dI/dz$ and $z_{s}$ is defined as the solution of the equation
\begin{equation}\label{eq:extremum}
A+NI^{\prime}\left(z_{s}\right)=0.
\end{equation}
Because we consider $N\gg1$, the important values of~$z$ are those for which $I(z)\ll1$, so  we can expand the logarithm and approximate
\begin{equation}\label{eq:interalI}
I\left(z\right)\approx\int\frac{dx}{\sqrt{2\pi}\sigma}e^{-\frac{x^{2}}{2\sigma^{2}}}\left[1-\exp\left(-ze^{-x}\right)\right].
\end{equation}
\begin{figure}
\includegraphics[width=0.45\textwidth]{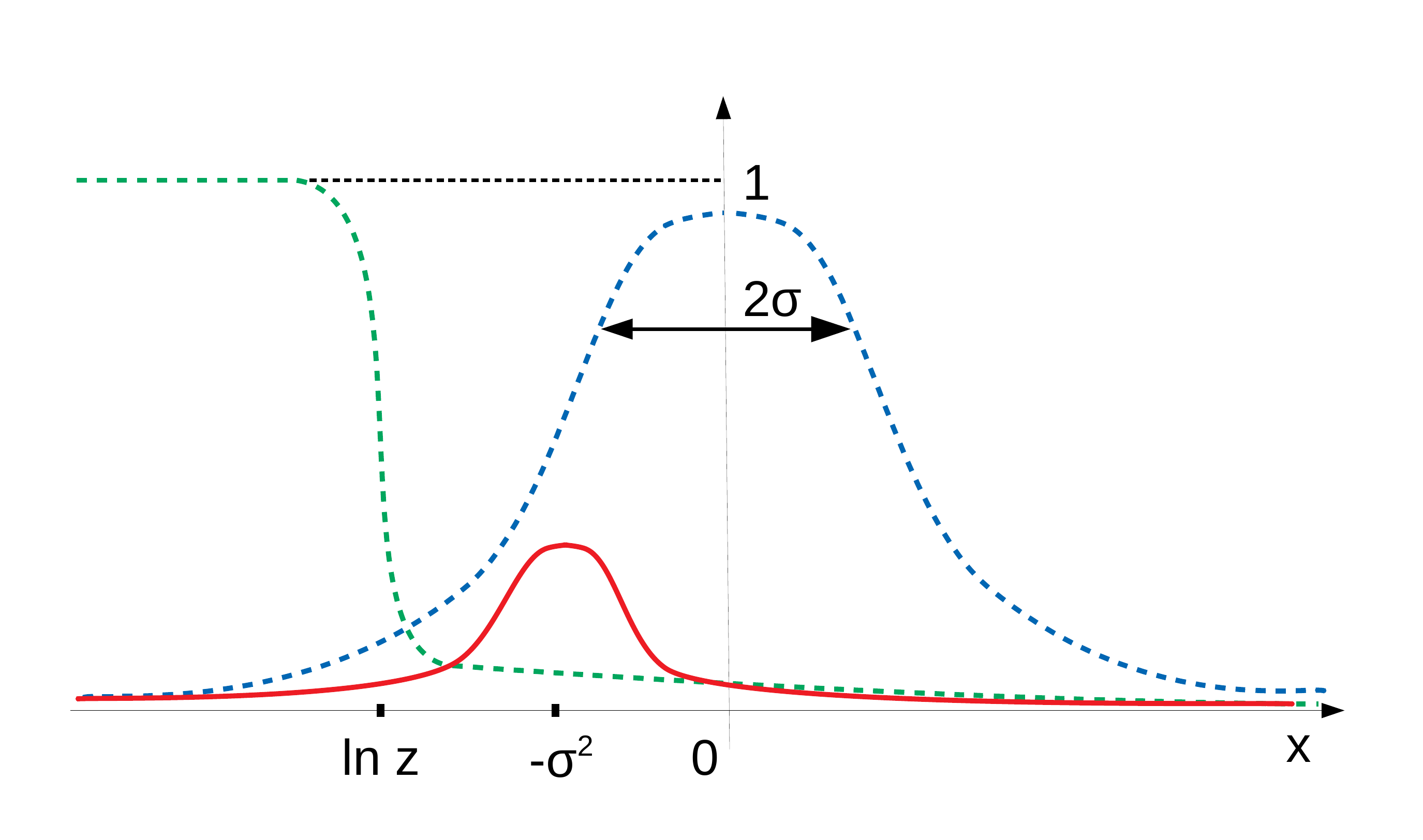}
\caption{\label{fig:Integral} Two factors of the integrand in $I(z)$ (\ref{eq:interalI}), $\exp(-\frac{x^{2}}{2\sigma^{2}})$ (dashed blue line) and $1-\exp\left(-ze^{-x}\right)$ (dashed green line), and their product (solid red line).}
\end{figure}
In the saddle point approximation the integral (\ref{eq:doubleintegral}) is determined by the small area near the real axis. To calculate $I(z)$ (\ref{eq:interalI}) we approximate $\exp\left(-ze^{-x}\right)\approx1-ze^{-x}$
for $x>\ln z$ and $\exp\left(-ze^{-x}\right)\approx1$ for $x<\ln z$. Then if $-\ln z\gtrsim\sigma^2+\sigma$ the integrand can be approximated as Gaussian for $x>\ln z$ and is suppressed for $x\lesssim \ln z$ \cite{Raikh1987} (see Fig.\ref{fig:Integral}):
\begin{align}
I\left(z\right)&\approx-\intop_{\ln z}^{\infty}ze^{\sigma^{2}/2}\frac{dx}{\sqrt{2\pi}\sigma}\exp\left[-\frac{\left(x+\sigma^{2}\right)^{2}}{2\sigma^{2}}\right]\nonumber\\
&=-\frac{ze^{\sigma^{2}/2}}{2}\mathrm{erfc}\left(\frac{\ln z+\sigma^{2}}{\sqrt{2}\sigma}\right),
\end{align}
Therefore, Eq.~(\ref{eq:extremum}) can be written as:
\begin{multline*}
A-N\frac{e^{\sigma^{2}/2}}{2}\mathrm{erfc}\left(\frac{\ln z_{s}+\sigma^{2}}{\sqrt{2}\sigma}\right)+\\
+N\frac{e^{\sigma^{2}/2}}{\sqrt{2\pi}\sigma}\exp\left[-\left(\frac{\ln z_{s}+\sigma^{2}}{\sqrt{2}\sigma}\right)^{2}\right]\approx0.
\end{multline*}
Introducing new variable $y=\frac{\ln z_{s}+\sigma^{2}}{\sqrt{2}\sigma}$ and considering $\left|y\right|\ll\sigma$ we obtain
\begin{equation}
y\approx\mathrm{erfc}^{-1}\left(\frac{2A}{Ne^{\sigma^{2}/2}}\right)-\frac{1}{\sqrt{2}\sigma}.
\end{equation}
Now we calculate the second derivative of $I$:
\begin{multline*}
I^{\prime\prime}\left(z_{s}\right)\approx\frac{e^{\sigma^{2}/2}}{\sqrt{2\pi}\sigma z_{s}}\exp\left[-\left(\frac{\ln z_{s}+\sigma^{2}}{\sqrt{2}\sigma}\right)^{2}\right]-\\
-\frac{e^{\sigma^{2}/2}}{\sqrt{2\pi}\sigma z_{s}}\frac{\ln z_{s}+\sigma^{2}}{\sigma^2}\exp\left[-\left(\frac{\ln z_{s}+\sigma^{2}}{\sqrt{2}\sigma}\right)^{2}\right]\approx\\ \approx\frac{\exp\left(3\sigma^{2}/2-\sqrt{2}\sigma y-y^{2}\right)}{\sqrt{2\pi}\sigma}.
\end{multline*}
resulting in
\begin{align}
f\left(A\right)\approx&\frac{\sigma}{Ne^{\sigma^{2}/2}}M^{1/2}\nonumber\\
&\times\exp\left(-Me^{\sqrt{2}\sigma Q-Q^{2}}+\frac{Q^{2}+\sqrt{2}\sigma Q}{2}\right),
\end{align}
where $Q=\mathrm{erfc}^{-1}\left(\frac{2A}{Ne^{\sigma^{2}/2}}\right)$
and $M=\frac{Ne^{-\sigma^{2}/2}}{\sqrt{2\pi}\sigma e}$.

For validity of the saddle-point approximation we need 
\begin{equation}
\left|NI'''(z_s)\left[\frac{1}{\sqrt{NI''(z_s)}}\right]^3\right|\ll1,
\end{equation}
where the quantity in the square brackets is the typical width of the relevant region near $z_s$. As a result, we obtain the condition
$N\gtrsim\sigma e^{\sigma^{2}/2-\sigma}$.

\end{document}